\documentclass[12pt,preprint]{aastex}

\shorttitle{NON-CIRCULAR MOTIONS}
\shortauthors{Mart\'{\i}nez-Garc\'{\i}a et al.}

\begin{document}

\title{EFFECTS OF NON-CIRCULAR MOTIONS ON
\\  AZIMUTHAL COLOR GRADIENTS}

\author{Eric E.\ Mart\'{\i}nez-Garc\'{\i}a}
\affil{Centro de Investigaciones de Astronom\'{\i}a, Apartado Postal 264,
       M\'erida 5101-A, Venezuela}
\email{emartinez@cida.ve}

\and

\author{Rosa A.\ Gonz\'alez-L\'opezlira and Gilberto C.\ G\'omez}
\affil{Centro de Radioastronom\'{\i}a y Astrof\'{\i}sica, UNAM, Campus Morelia, Michoac\'an, M\'exico, C.P. 58089}
\email{r.gonzalez@crya.unam.mx, g.gomez@crya.unam.mx}

\begin{abstract}
Assuming that density waves trigger star formation, and that young stars preserve the
velocity components of the molecular gas where they are born, we analyze the effects that
non-circular gas orbits have on color gradients across spiral arms. We try two
approaches, one involving semi-analytical solutions for spiral shocks, and another
with magnetohydrodynamic (MHD) numerical simulation data. 
We find that, if non-circular motions are ignored, the comparison 
between observed color gradients and 
stellar population synthesis models would in principle yield pattern speed values that are
systematically too high for regions inside corotation, with the difference
between the real and the measured pattern speeds increasing with decreasing
radius. On the other hand, image processing and pixel averaging
result in systematically
lower measured spiral pattern speed values, regardless of the kinematics
of stellar orbits. The net effect is that roughly the correct pattern
speeds are recovered, although the trend of higher measured $\Omega_p$ 
at lower radii (as expected
when non-circular motions exist but are neglected) should still be observed.
We examine the \citet{mar09} photometric data and confirm 
that this is indeed the case.
The comparison of the size of the systematic pattern speed offset in
the data with the predictions of the semi-analytical and MHD models 
corroborates that spirals 
are more likely to end at Outer Lindblad Resonance, as these authors had already found. 

\end{abstract}

\keywords{ galaxies: kinematics and dynamics --- galaxies: magnetohydrodynamic simulations --- galaxies: stellar content
--- galaxies: spirals --- galaxies: structure}

\section{Introduction.}

Until the detection of an azimuthal color gradient across one of the arms 
of the SAc galaxy M~99 \citep[][GG96 hereafter]{gon96}, only sparse evidence
of star formation triggered by spiral density waves had been found
in stellar counts in the Milky Way\citep{sit89,sit91,ave89} and M~31\citep{efr80a,efr80b,efr85}.
More recently,
applying the same method defined by GG96 and described below, \citet[][MG09 hereafter]{mar09} 
examined a sample of 13 spiral galaxies 
of types A and AB, and found color gradients consistent with theoretical expectations in 10 of their
objects.\footnote{See also \citet{gro09} for an infrared study of young stellar
complexes in NGC~2997.} 
Although they did not compute stellar orbits, both GG96 and MG09 
implicitly assumed that all stars, including those recently born 
near spiral arms, move in circular orbits. 
MG09 did investigate the effects of variable circular speeds and variable densities
on observed color gradients and found them to be negligible. 
However, complicated non-circular motions have been reported in studies about 
the migration of young stars following star formation triggered 
by spiral shocks \citep{yua69,wie79,fer08}.
\citet{bash81} had also noticed that the ballistic orbits calculated from galactic HII
region complexes follow non-circular trajectories that initially move along
(and not across) the spiral arms. 

In the present work, we 
examine how the presence of non-circular motions would modify the 
pattern speeds derived from color gradients under the
assumption of circular orbits.

We will explore two different approaches.
The first one (see \S~\ref{appr1}) involves the semi-anaylitical solutions obtained 
for spiral shocks \citep{rob69,shu73,gitt04};
we assume that newly born stars preserve the orbital motion of the shocked gas where they form.
The second approach (see \S~\ref{appr2}) is based on data from MHD simulations;
we follow the gas flow vectors near spiral arms. We always assume that stars are
triggered all along the studied regions of spiral arms in the respective model.
Other methods, not discussed here, may involve orbit calculations for young 
stellar groups.

\section{The GG96 method.} \label{themethod}

The photometric technique employed by GG96 and MG09 uses 
three optical bands, $g$, $r$ and $i$, plus the near
infrared $J$ band (see table \ref{tbl-filters}). 
With these filters, the reddening insensitive and star-formation 
sensitive photometric index $Q$ is obtained: 

\begin{equation}
  Q(rJgi) = (r-J) - 0.82 (g-i),
\end{equation}
\begin{equation} \label{eqQlog}
    Q(rJgi) = \log_{10} \frac{I^{2.05}_{g} I^{2.50}_{J}}{I^{2.50}_{r} I^{2.05}_{i}}.
\end{equation}

\noindent
In star-forming regions, the $Q(rJgi)$ index has higher values, because the \emph{g} and \emph{J} 
bands in the numerator of equation \ref{eqQlog} trace the light from 
red and blue supergiants, respectively. When the star-formation activity is poor, the 
value of the $Q(rJgi)$ goes down. 

\begin{deluxetable}{ccc}
\tabletypesize{\scriptsize}
\tablecaption{Filters employed in the GG96 method\label{tbl-filters}}
\tablewidth{0pt}
\tablehead{
\colhead{Filter} & \colhead{$\lambda_{eff}$} & \colhead{FWHM}
}
\startdata
\emph{$g$}       & 5000\AA & 830\AA \\
\emph{$r$}       & 6800\AA & 1330\AA \\
\emph{$i$}       & 7800\AA & 1420\AA \\
\emph{$J$}       & 1.25\micron & 0.29\micron \\
\enddata
\end{deluxetable}

\begin{figure}
\epsscale{1.00}
\plotone{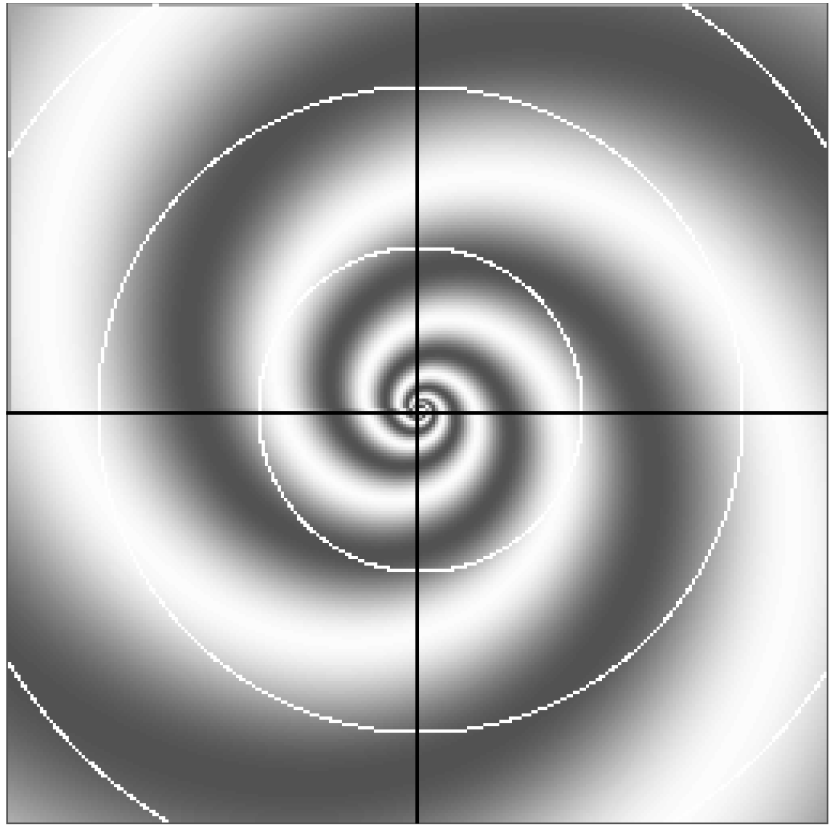}
\caption{Schematic of a spiral disk.
White circles are separated by 50 arbitrary units.  \label{spiral}}
\end{figure}

\begin{figure}
\epsscale{1.00}
\plotone{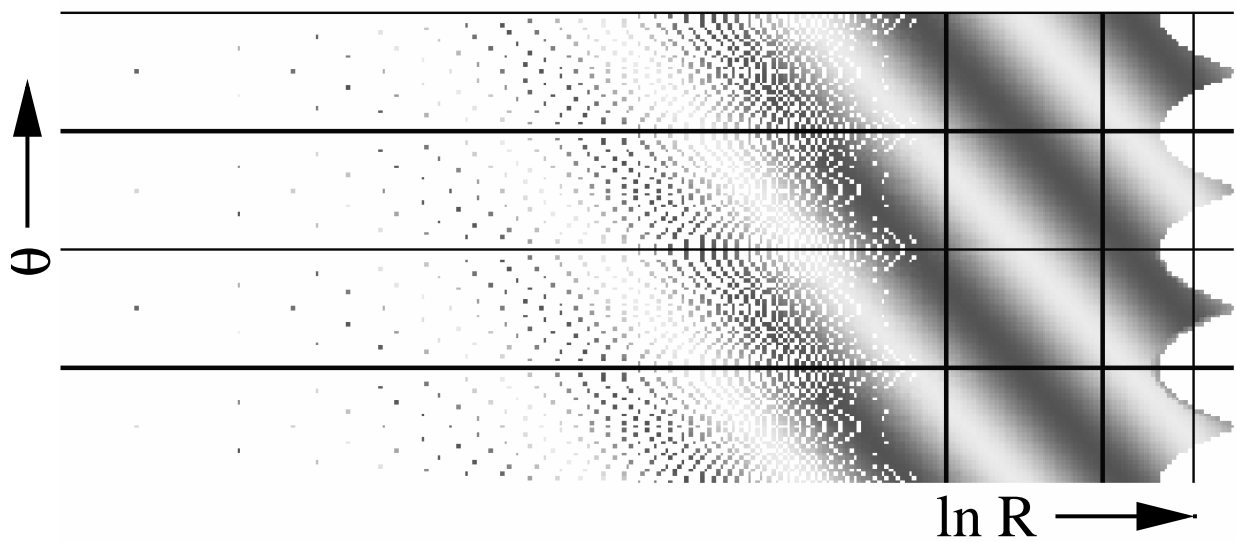}
\caption{Unwrapped version of figure \ref{spiral}. Logarithmic spiral
arms now appear as straight lines with slope = cot(-$i$), where $i$
is the arm pitch angle. Horizontal lines are located every $90 \degr$.
Vertical lines correspond to radii marked 
with white circles in figure \ref{spiral}. \label{unwra}}
\end{figure}

\begin{figure}
\epsscale{1.00}
\plotone{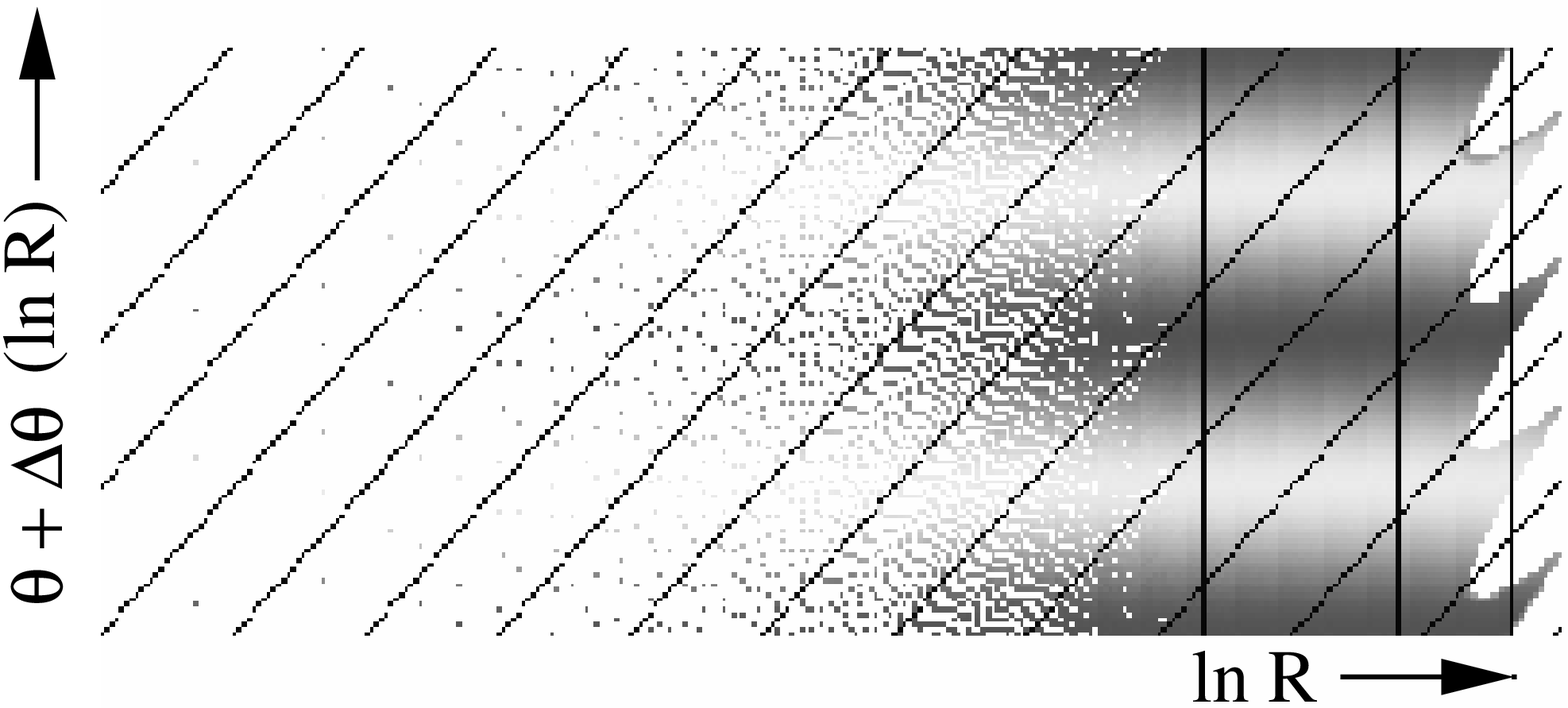}
\caption{Straightened version of figure \ref{unwra}. Now spiral
arms appear as horizontal lines. The $\ln R$ coordinate remains the same,
but the $\theta$ coordinate has a phase-shift that depends on $\ln R$.  \label{strai}}
\end{figure}

For each studied galaxy and at each wavelength,
the spiral arms are first unwrapped \citep{iye82},
by plotting them in a $\ln R$ versus $\theta$ map (see figures \ref{spiral}, \ref{unwra}). 
All the spiral arms in the disk galaxy are inspected, 
in search of color gradients. If a candidate region is found, 
the arm is straightened, by adding a different phase-shift to $\theta$ at 
each value of $\ln R$, until the arm appears as a horizontal line
(see figure \ref{strai}). This procedure allows us to add the data from different $\ln R$, and thus
to increase the signal-to-noise ratio of the light profile as a function of $\theta$.
1-D plots of $Q$ vs.\ $\theta$ are obtained from the unwrapped and straightened galaxy images.
These are then compared to 1-D plots of $(g-J)$, that 
trace the dust lane location, and to 1-D plots of the $K$-band (or $K^\prime$ or $K_{s}$, 
depending on what we have available) data,
that trace the density wave (under the assumption
that the near-IR emission of red supergiants can be neglected).
The regions that seem to match theoretical expectations according to these plots
are selected and compared to stellar population synthesis models, that provide 
$Q$ as a function of stellar age, $t_{\mathrm age}$.

If one assumes that stars form in the site of the shock, and that they age as they move away 
from this location, then distance from the dust lane (at constant radius) parametrizes 
stellar age. In fact, stretching the model $Q(t_{\mathrm age})$ to fit the data 
(where $Q$ is a function of angular or linear distance) fixes the ratio between the 
distance from the shock, $d$, and the age of the stellar population. If, in addition, the rotational 
velocity is known, it is possible to find the angular velocity of the spiral pattern,
$\Omega_{p}$, as follows:

\begin{equation} \label{eqOMEGA}
  \Omega_{p} \cong \frac{1}{R_{\mathrm mean}} \left(v_{\rm rot}(R) - \frac{d}{t_{\rm age}} \right).
\end{equation}

\noindent Here, $R_{\mathrm mean}$ is the mean radius of the region where the gradient has been detected, 
measured from the center of the galaxy; $v_{\mathrm rot}(R)$ is the rotational velocity of the disk
at such radius (obtained from the literature); $d$ is distance from the shock,
and $t_{\mathrm age}$ is the stellar age obtained from fitting the population synthesis model to the observations.
With the pattern speed, we obtain the location of major resonances:
the Outer Lindblad Resonance (OLR), corotation, and the 4/1 resonance. 
This resonance positions are compared with the observed spiral end points.

As mentioned before, this procedure implicitly assumes circular motion for the stars involved,
even though no stellar orbits have been computed.

\section{Semi-analytical shock solutions.} \label{appr1}

\begin{figure}
\epsscale{1.00}
\plotone{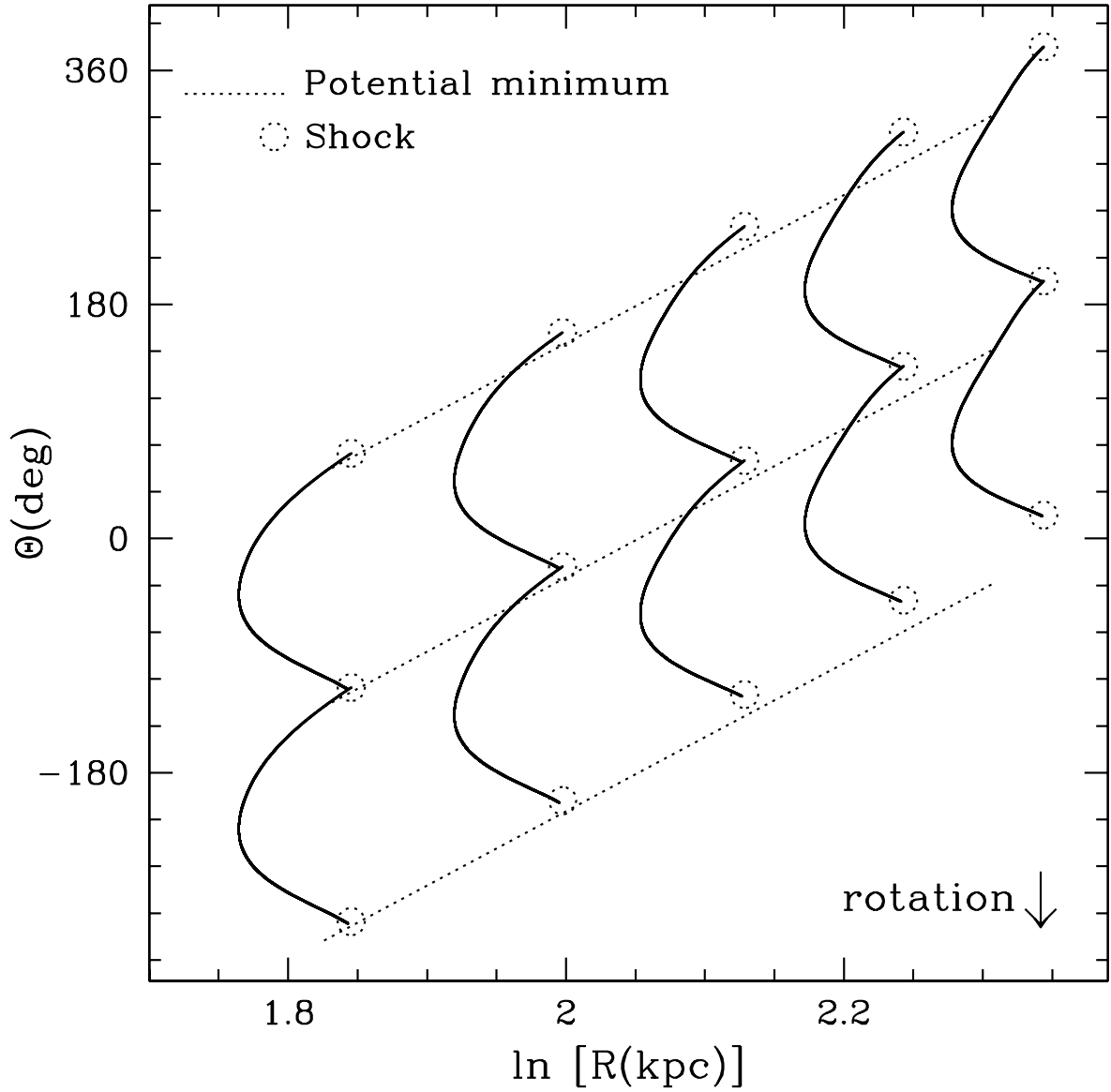}
\caption{Typical streamlines obtained from semi-analytical solutions of spiral, trailing type, shocks.
{\it Dotted line:} spiral potential minimum;
{\it dotted circles:} shock position.
Average radii, from left to right, are 6, 7, 8, 9, and 10 kpc.
\label{sharp}}
\end{figure}

Shock solutions obtained from semi-analytical approaches
have shown that gas streamlines are not circular. 
\citet{rob69}, for example, found that streamlines appear as sharp-pointed ovals
(see figure~\ref{sharp}). 
Although such 
solutions were found for the tightly wound approximation, they provide
a good reference for the study of real galaxies~\citep[see also][]{saaf74}.
Here, we obtain gas streamlines under the assumption that young stars follow the motion of the
molecular gas, and by adopting a procedure similar to that of \citet{gitt04}.\footnote{ 
See appendix~\ref{appsemianalyt}. The reference system is the rotating frame of the spiral
pattern.} 
Our adopted model has a pattern speed of 13 km s$^{-1}$ kpc$^{-1}$, and a flat rotation
curve with $v_{\mathrm rot}=220$ km s$^{-1}$.
From $R=5$ kpc to $R=10$ kpc, streamlines were calculated in steps of 0.05 kpc.
Since the stellar models give us $Q$ as a function of $t_{age}$, we can assign
a stellar age to each point in the streamlines;
in order to do so, we need to assume that the onset of star formation 
occurs at a certain orbital
time $t$, where $t=0$ corresponds to the shock position.

\subsection{Star formation onset delay.} \label{appr1_sub}

\citet{mar09} use the term ``timescale for star formation" as 
the duration of the burst, that is estimated to be $~2 \times 10 ^ 7$ years.
They do not make any assumptions about the time it takes for star formation to begin 
after the shock. A preliminary inspection of their findings shows
that the shock (as traced by the dust lanes) and the
onset of star formation (marked by $t_{age}=0$ in the stellar models)
are not always at the same location.
In real galaxies, whether there is a delay between the shock and the star formation onset
is not well understood yet. In fact, the time needed for a diffuse cloud
of neutral gas to first become a dense cloud, then a molecular cloud,
and finally a self-gravitating cloud is $\sim10^7$ years. Once
this cloud is formed, the onset of star formation may be very fast \citep{vaz07,hei08}.
\citet{egu09}, 
based on an observational study between the peaks of molecular ($CO$) and young
stellar ($H_{\alpha}$) arms, report a delay for the onset of star formation of $\sim5-30$
Myr in 5 galaxies (out of 13); \citet{tam08} estimate a timescale for star formation in the range 1-4 Myr,
from a study of angular offsets between HI and 24$\mu m$ emissivity
peaks in a sample of 14 disk galaxies.\footnote{For these authors, the term 
``star formation timescale" is equivalent to star formation onset delay.} 

In order to test for the effects of a star formation onset delay on
azimuthal color gradients, we try three cases. The first one has a delay 
(i.e., the time it takes for the gas to move from the shock to the location 
where star formation begins)
$\delta t_{shock}=0$ yr;  
the other two have $\delta t_{shock}=1\times10^7$ yr and 
$\delta t_{shock}=2\times10^7$ yr, respectively.

\section{MHD simulations.} \label{appr2}

The gas simulations were performed with a version of the {\sc zeus} 
code \citep{sto92a,sto92b}, that is an eulerian, time-explicit,
finite-difference code for ideal MHD simulations.
We employ a 2D grid with 500 $\times$ 500 points
in polar coordinates. The radial extent goes from $R \sim$ 3 to 30 kpc,
and the azimuthal one from $\theta \sim 0$ to $\pi$ radians. We assume
that the half disk simulation data have a $180 \degr$ ``mirror'' symmetry.
The gaseous disk follows an isothermal equation of state with a
temperature of 10900$\degr$ K.
At the beginning of the simulation, the magnetic field has a
toroidal geometry and a value of $5\,\mu$G at $8.5$\,kpc from the
galactic center, although it rapidly evolves away from this setup.
No self-gravity is included.

The background gravitational potential has two components: one axisymmetric,  
that consists of a bulge, a disk, and a halo, while the other is the
non-axisymmetric spiral arm perturbation.
The adopted bulge and disk are described in \citet{all91}.
The dark matter halo is NFW type \citep{nav96,nav97}, with
density:


\begin{equation}
  \rho(R) = \frac {\rho_{h}} { {R}/{a_{h}} \left( 1 + {R}/{a_{h}} \right)^{2}},
\end{equation}

\noindent where $\rho_{h}$=$1.021 \times 10^{-2}$ $M_{\sun}/pc^{3}$
and $a_{h}$=15.133 kpc.
The two-arm spiral potential has a pitch angle of $15.5\degr$, and
is self-consistent in the stellar orbits sense \citep{pic03}.
The simulation is performed in this stellar arms reference frame, that 
rotates with velocity $\Omega_p = 20$~km\,s$^{-1}$\,kpc$^{-1}$.

\begin{figure}
\epsscale{1.00}
\plotone{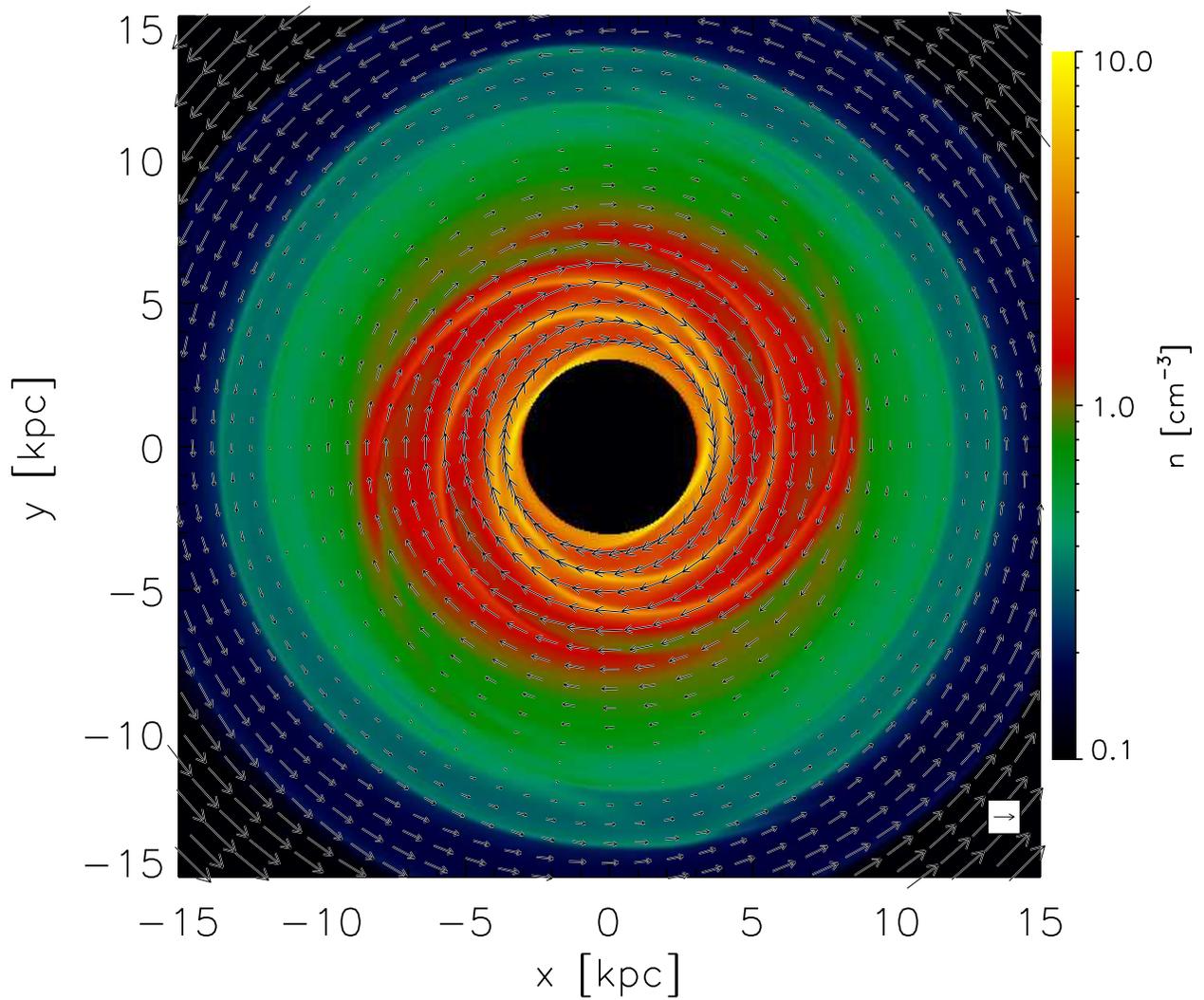}
\caption{MHD simulation at 800 Myr.
Colors represent the gas density, while
arrows indicate the velocity field (in the rotating frame of 
the spiral pattern).
The arrow in the lower right corner corresponds to $\sim$ 80 km s$^{-1}$.
Corotation is located at $R=10.9$ kpc.
\label{gilgosim}}
\end{figure}

We allow the simulation to evolve during $8 \times 10^{8}$\,yr
(see figure~\ref{gilgosim}).
By this time, the gas develops spiral arms as a response to
the imposed perturbation,
initial transients subside, and changes in the simulation are 
only observable at very long time scales.
Therefore, we assume that density and velocity distributions at this
time represent a close approximation to a steady state solution.

As shown in figure~\ref{gilgosim}, the gas in the simulation
has responded to the spiral perturbation with a four arm pattern \citep{marto04}.
A hypothetical circular gas orbit would encounter
four shocks before completing its transit around the disk.
In this case, the corresponding velocity field
would be very different from the case where only two shocks are considered.
For the present investigation we focus on the main shock
(i.e., the one closer to the potential minimum of the spiral
perturbation), although spiral-arm triggering of star formation
may also take place in the secondary shocks.

\begin{figure}
\epsscale{1.00}
\plotone{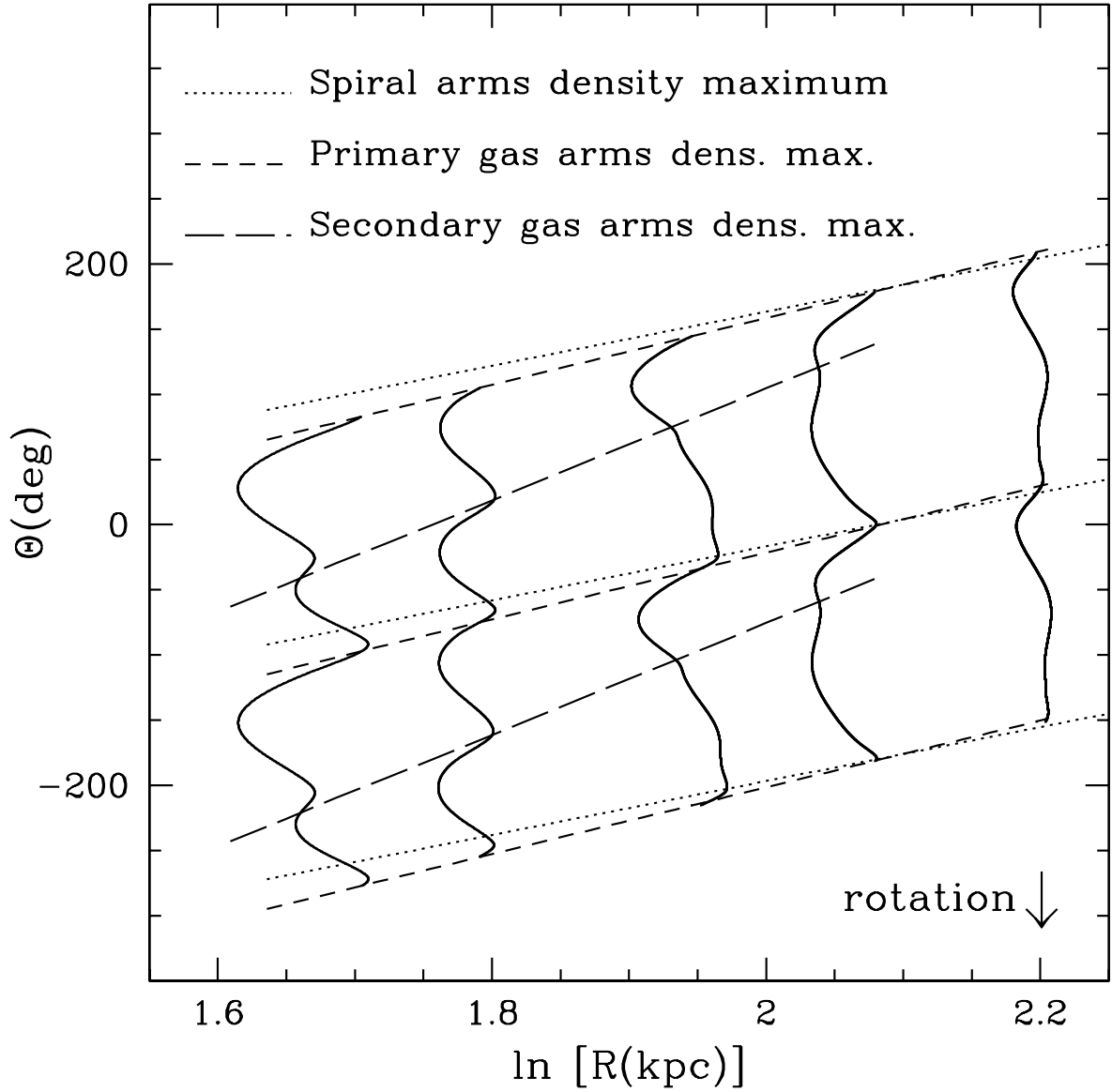}
\caption{Gas orbits for the simulation data, in the $\ln{R}$ vs.\ $\Theta$ plane.
Initial radii (from left to right) are 5.5, 6, 7, 8, and 9 kpc.
{\it Dotted line:} stellar arms density maximum;
{\it short-dashed line:} primary gas arms density maximum;
{\it long-dashed line:} secondary gas arms density maximum. \label{gorbits}}
\end{figure}

In order to obtain the orbit a parcel of gas would follow, we locate
the main shock position by searching for the gas density maxima
closest to the (stellar) spiral arms.
These are taken as the start point for the integration of the gaseous
orbits.
The orbits are then calculated from the velocity data (in the
spiral pattern reference frame), using a Runge-Kutta method, 
every $\Delta R = 0.005$\,kpc, and $\Delta t=7\times10^{12}$\,s.
A sample of orbits are shown in figure~\ref{gorbits}.
Other gas orbits with MHD simulations using the full time evolution of the gas
can be seen in~\citet{gom09}, who find a close resemblance between gas orbits
and the central family of ballistic orbits in stellar dynamics.

\section{Analysis and Results.}

\subsection{$Q$ profiles.}

\begin{figure}
\epsscale{1.00}
\plotone{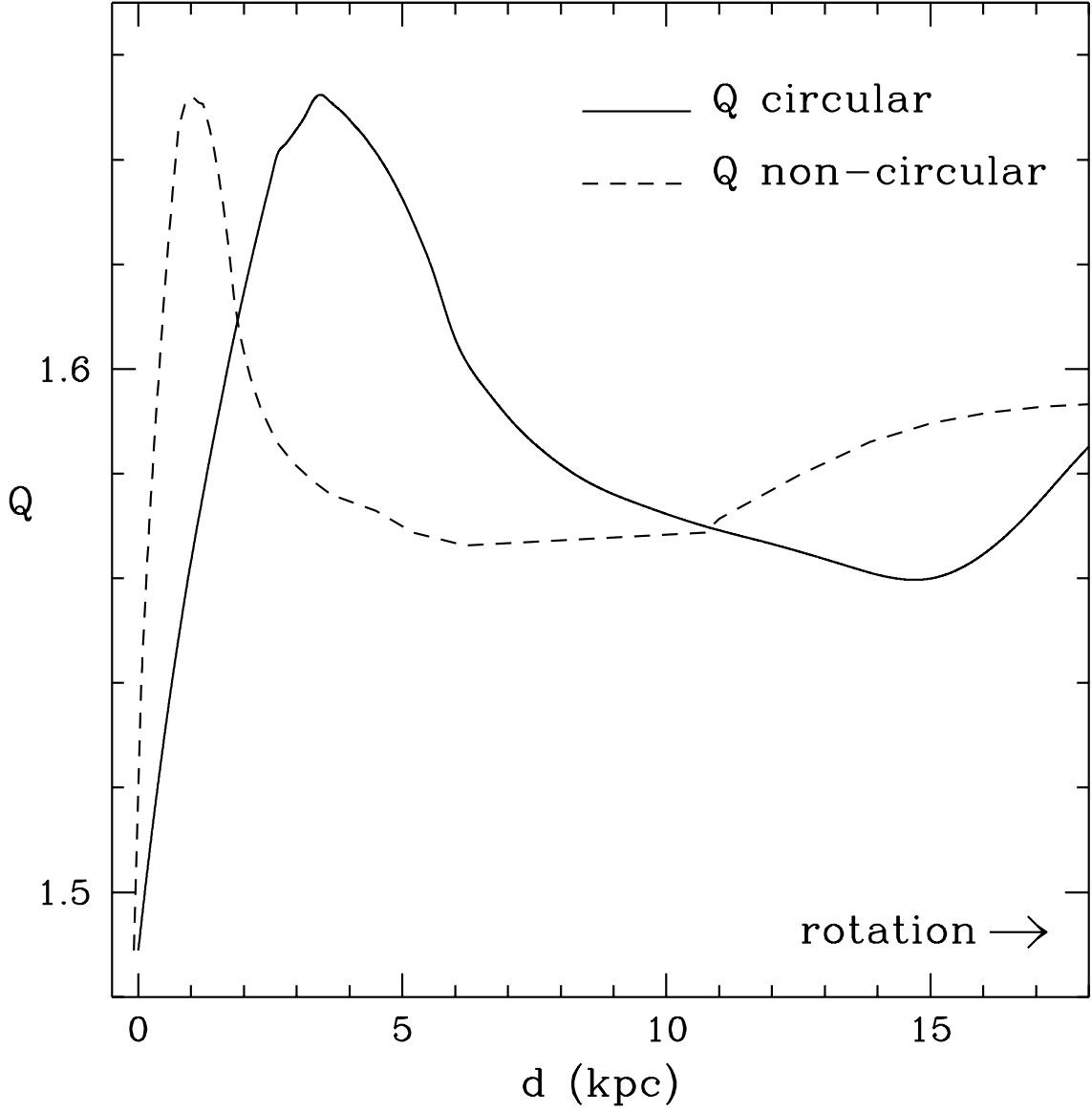}
\caption{$Q$ index profiles inside a narrow annulus with mean radius $R=7$ kpc,
vs.\ azimuthal distance relative to the shock ($d$=0).
{\it Solid line:} circular case; {\it dashed line:} non-circular case. \label{diff_Q}}
\end{figure}

From the first approach, involving semi-analytical solutions, we obtain 
$Q$ index profiles for the non-circular and circular cases,
respectively, calculated in a narrow annulus with mean radius $\sim 7$ kpc;
we show them in figure~\ref{diff_Q}.
To produce the $Q$ profiles, we assume a background population of old stars with an age of $5\times10^9$ yr
(i.e., the density wave),
plus a young burst of star formation lasting $2\times10^7$ yr. 
Both the old and the young populations have a Salpeter IMF, with $M_{lower}=0.1 M_{\sun}$
and $M_{upper}=10  M_{\sun}$.\footnote{
There is a conspicuous inverse correlation between the detection of
azimuthal color gradients and the presence of HII regions (see MG09).}
For the present case, we adopt a constant fraction of young
stars of 2\%~by mass (see equation 9 in MG09).\footnote{
The effects of variable densities of both the young and old populations on the
color gradients have already been
discussed by \citet{mar09}. Their main conclusion is that variable stellar densities
can produced deformations in the expected color gradients. However, the estimated error 
introduced by these deformations is lower than the computed random error contributed by 
the combined uncertainties in the inclination angle, the rotation velocity, and the
distance to the galaxy.} We then use the stellar population synthesis models by 
\cite{cha07} to find the changing $g$, $r$, $i$, and $J$ emission of the population mixture
as it evolves. 

\begin{figure}
\epsscale{1.00}
\plotone{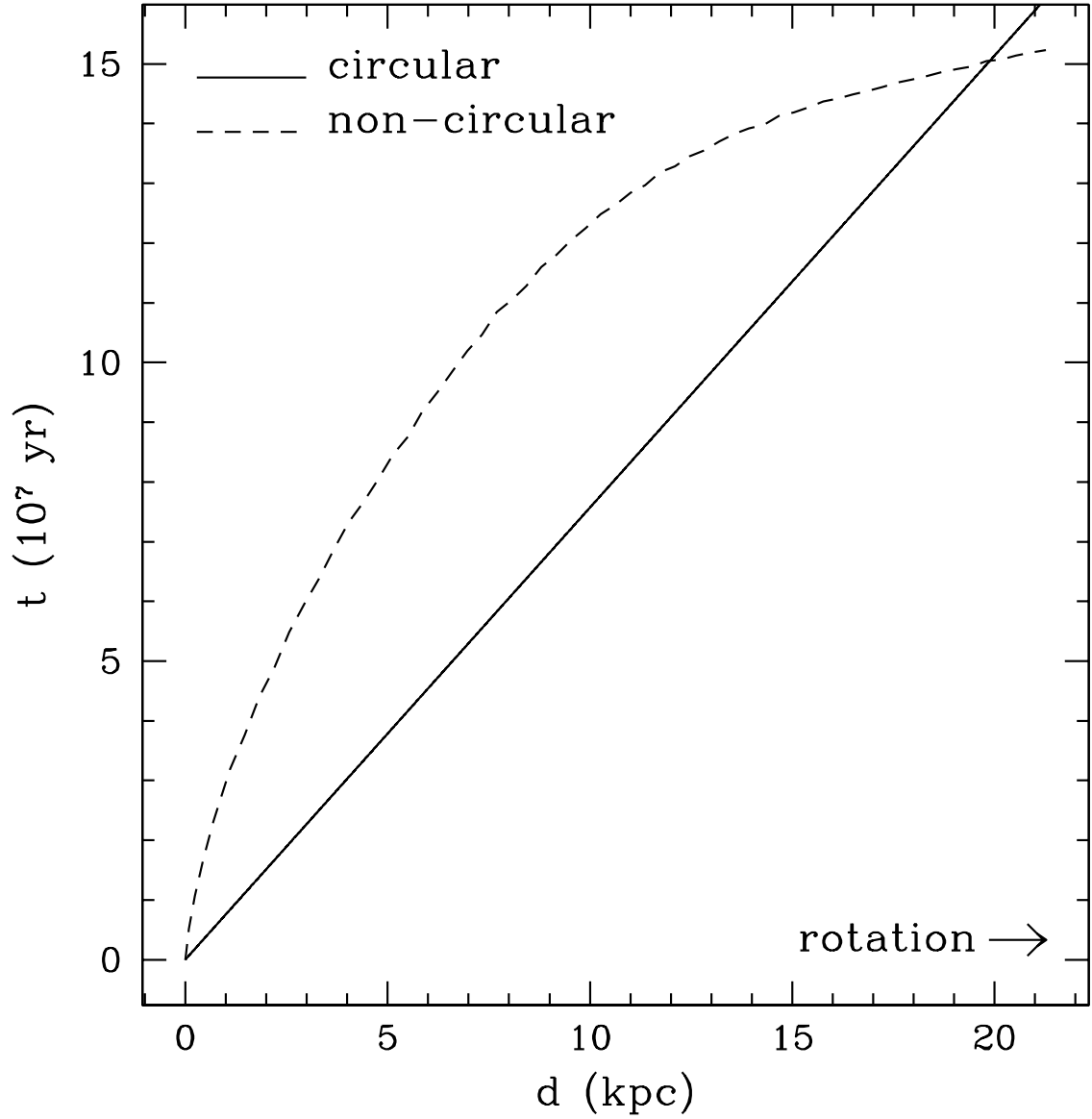}
\caption{Stellar population ages inside a narrow annulus with mean radius $R=7$ kpc, vs.\
azimuthal distance from the shock ($d$=0).
{\it Solid line:} circular case; {\it dashed line:} non-circular case.  \label{diff_t}}
\end{figure}

In the circular case, the stellar complexes
that give origin to the $Q$ profile 
were all born at the same ($\ln R, \theta$) position. In the non-circular case, due to the 
trajectories of the gas streamlines,
the stars that give rise to the $Q$ profiles were not all formed at the same location. 
The age of the stellar complexes that contribute to the $Q$ profile as a function of
distance from the shock position is plotted in
figure~\ref{diff_t}. As expected, young objects tend to spend more time concentrated toward 
the spiral shock ($d=0$ kpc), with the consequence that the peak of the $Q$ profile
({\it dashed line} in figure~\ref{diff_Q}) occurs closer to the shock.

Another aspect to be noticed in figure~\ref{diff_Q} is that there is no ``downstream fall"
of the gradients (i.e., $Q$ does not fall below, in this case,
$\sim 1.56$, contrariwise to what is sometimes observed with the data). 
\citet{mar09} had hypothesized that the fall of the observed $Q$ 
profiles below the model values (assuming pure circular orbits) might be caused 
by stellar non-circular motions in the data. 
However, results may differ if ballistic trajectories with postshock
velocities are considered~\citep[e.g.][]{bash77,bash79}.

\subsection{Pattern speeds.}

From the $Q$ profiles, it is possible to derive pattern speeds 
by comparing stellar population synthesis models ($Q$ vs.\ $t_{\mathrm age}$) to the ``data"
($Q$ vs.\ $d$), assuming implicitly that 
stars move in circular orbits, as was done by GG96 and MG09, and was sketched above in section \ref{themethod}. 

\begin{figure}
\epsscale{1.00}
\plotone{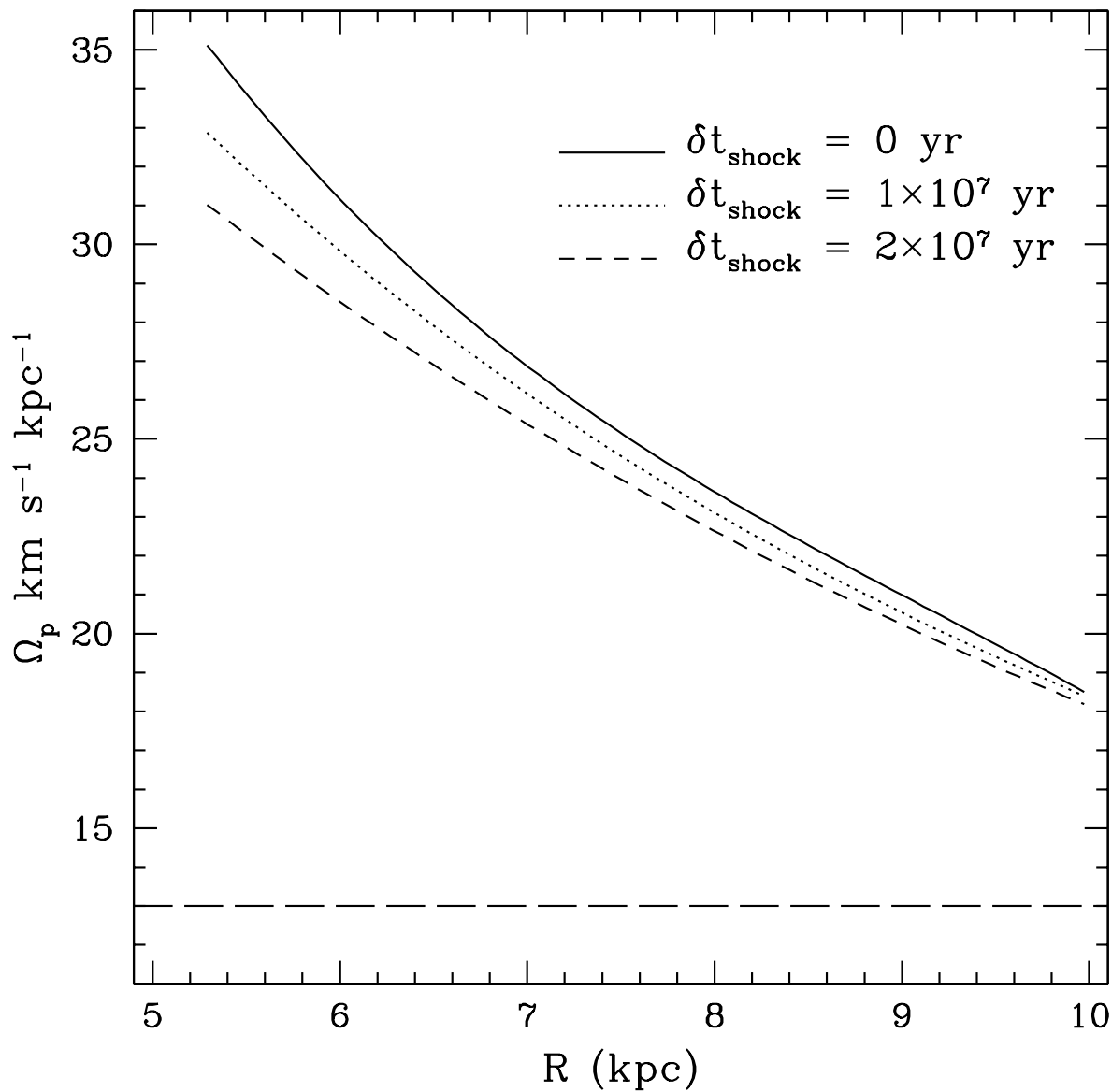}
\caption{$\Omega_{p}$ values obtained at different galactocentric radii, from the numerical semi-analytical solutions
of spiral shocks, under the (false) assumption that stars move in circular orbits.
{\it Solid line:} star formation onset delay $\delta t_{shock}=0$ yr;
{\it dotted line:} $\delta t_{shock}=10$ Myr;
{\it short-dashed line:} $\delta t_{shock}=20$ Myr. 
{\it Long-dashed line:} spiral perturbation input angular velocity. \label{delayt}}
\end{figure}

In the case of the semi-analytical approach,
``data" $Q$ profiles were obtained from narrow annular regions, $\sim 0.001$ kpc wide. 
Given that we know the input pattern speed, we can search for systematic effects
in the determination of $\Omega_{p}$. Figure~\ref{delayt} shows the input pattern speed 
({\it long-dashed line}), and 
$\Omega_p$ values that are obtained from the color gradients 
for three different presuppositions about the delay of star formation onset after the shock
(0, 10, and 20 Myr). 
For reasons explained below in~\S~\ref{lnRcoll}, rather than stretching the stellar population synthesis 
model to the ``data", we just compare the  
positions of the maxima of the model $Q$, on the one hand, 
and of the shock location, on the other. 

There is a significant systematic effect, whereby the derived $\Omega_p$ is always larger than
the input pattern speed; moreover, the effect 
decreases with galactocentric radius. Also, the 
difference between the input and the output pattern speeds decreases for larger 
star formation onset delays.

\begin{figure}
\epsscale{1.00}
\plotone{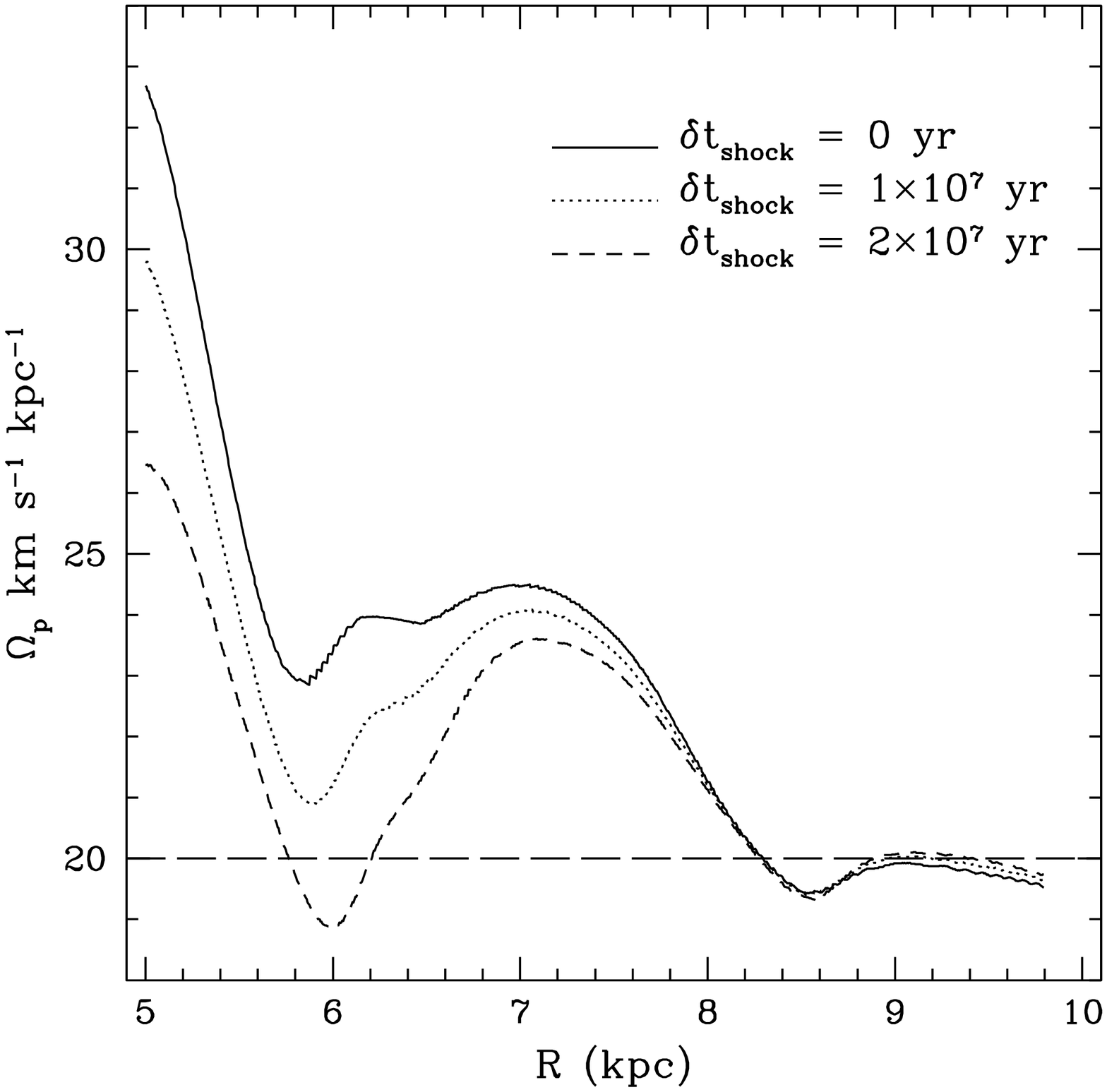}
\caption{$\Omega_{p}$ values obtained from the MHD simulation data, at different galactocentric radii,
assuming a circular motion dynamic model.
The ``data" $Q$ profiles were calculated in narrow annuli 0.001 kpc wide.
Symbols as in figure~\ref{delayt}. \label{apcm}}
\end{figure}

In the case of the MHD simulations, we again analyze the pattern speeds 
derived when assuming a circular motion dynamic model. 
In order to obtain $\Omega_p$, we compare the peak value of $Q$ in the stellar
population model to the density maximum of the gas in the simulations.
In figure~\ref{apcm}, we show the $\Omega_{p}$ values
obtained for $\delta t_{shock}$ of 0, 10, and 20 Myr. 

\section{Pixel averaging} \label{lnRcoll}

One of the caveats of the GG96 method, that was not discussed in MG09, comes
from image processing. Each pixel in the optical and infrared
images actually contains information about many young stellar orbits that all
fall within the same spatial region. 
The process of ``unwrapping" the spiral arms then averages pixels in the
$\theta$ and $\ln{R}$ directions (the ``straightening" of the arms is just a shift
in the $\theta$ direction and does not involve pixel averaging). 
In order to increase the signal-to-noise ratio of the $Q$ profiles, an additional averaging in the $\ln{R}$ direction 
is done in selected regions. 
This last step averages together orbits with different angular speeds, and hence stars 
of slightly different ages at a fixed distance from the shock.


Figure~\ref{cavfig} shows the behavior of $Q$ profiles
from an unwrapped and straightened synthetic image (see below),
under the assumption of pure circular motion; the different profiles are 
taken from various mean radii, after averaging in $\ln{R}$.
The shift of the peak toward $d=0$ is expected as we approach the corotation radius.
However, due to the pixel averaging, a drop in the maximum value of $Q$ is also obtained.
If, in order to match the 
observed or synthetic profile peak, a constant downward shift is applied to the stellar population synthesis model, 
\footnote{Vertical shifts are also expected if the metallicities, or the 
ratios of young to old stars of model and data are different; these differences do not
significantly affect the derived pattern speed (MG09),
because the horizontal extension of the profiles is not noticeably affected.}
the ``wings'' of the profiles will not match,
since those of the stellar model will now lie within the ``wings'' of the data.
This mimics wider gradients, and 
in order to fit the observed or synthetic profile we will need to 
overstretch the model.
Consequently, the pattern speed obtained
via equation~\ref{eqOMEGA} 
will be lower, because  
the stellar age ($t_{\mathrm age}$) 
at a fixed distance $d$ from the shock will be underestimated. 
The effect amounts to $\sim$ 1 to 2 km s$^{-1}$ kpc$^{-1}$ 
for moderate pixel averaging, but can be as large as
10 km s$^{-1}$ kpc$^{-1}$ when the averaged region in $\ln{R}$ is very extended.

One way out of this problem is to compare 
only the distances between the onset of star formation and the $Q$ peak, in the model and the data, 
instead of the whole $Q$ profiles,
since image processing affects less the peak positions than the profile shapes.
We have used this approach with our semi-analytical calculations and MHD simulations; unfortunately,
in the case of real data, the position of the star formation onset
is generally unknown (see also~\S~\ref{appr1_sub}).

\begin{figure}
\epsscale{1.00}
\plotone{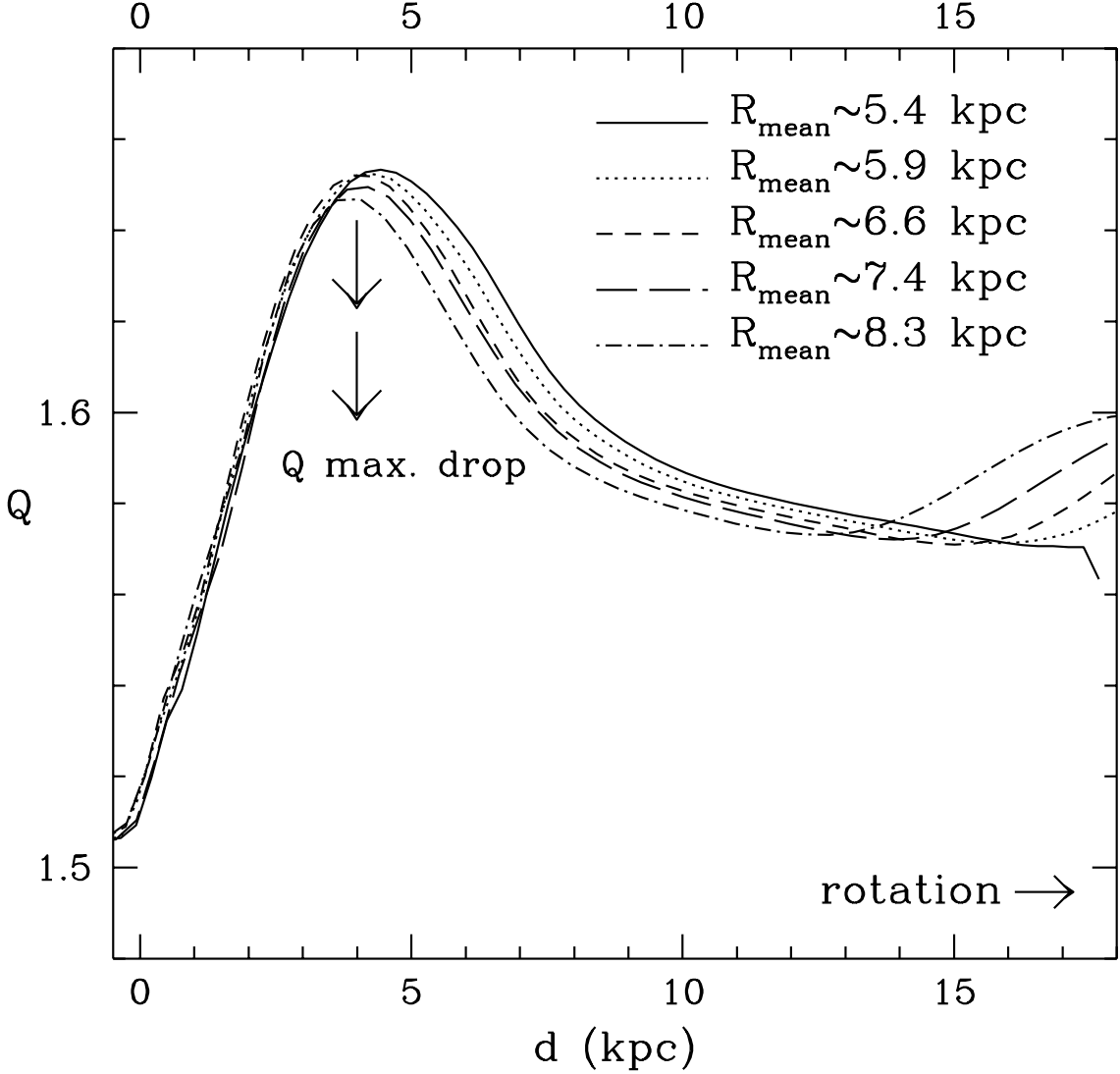}
\caption{$Q$ profiles obtained from unwrapped and straightened images at different 
mean radii, after averaging in $\ln{R}$. The dynamic model assumes pure circular motions with
$v \sim$ 220 km s$^{-1}$, and $\Omega_{p}$= 13 km s$^{-1}$ kpc$^{-1}$.
$R_{\mathrm{mean}}$ denotes the region mean radius, $d$ is the azimuthal distance
from the shock ($d=0$). Stellar population synthesis models from \citet{cha07},
with IMF lower limit of 0.1 $M_{\sun}$, and upper limit of 10 $M_{\sun}$.} \label{cavfig}
\end{figure}

To better estimate the effects of averaging,
with the $R$, $\Theta$ and $t$ (time elapsed since the shock, $t$=0 corresponds to the shock location) 
data of the streamlines obtained from the semi-analytical solutions,
we constructed images of a synthetic ``galaxy" as it would
be observed in the $g$, $r$, $i$ and $J$ bands
with the aid of stellar population synthesis models \citep{cha07} 
(the fraction of young stars constitute is $2\%$ by mass).
The model face-on galaxy is at a distance of 35 Mpc and the synthetic images
have $1\arcsec \times 1\arcsec$ pixels.
Whenever various streamlines were located at the same pixel, their 
elapsed times since the shock were averaged. 
Next, the synthetic images were unwrapped. Each hyperpixel of the unwrapped images is $3\degr$ long in the 
$\theta$ axis; the $\ln{R}$ axis has 250 hyperpixels in total, from the
center of the galaxy to $R=10$ kpc. 
Then, the synthetic unwrapped images were straightened according to the shock's pitch angle
($\sim 5.266 \degr$), that is smaller than the potential minimum's pitch
angle ($\sim 5.739 \degr$; see appendix ~\ref{appsemianalyt}).
Finally, $Q$ profiles were obtained from these ``images" in sections with a width of  
5 hypercolumns in $\ln{R}$.
Pattern speeds were measured by comparing the ``observed" $Q$ (vs.\ $d$) from the 
simulated images, at different radii, with  $Q$ (vs.\ $t_{\mathrm age}$) 
from stellar population synthesis 
models.\footnote{The adopted models throughout this investigation
have an IMF upper limit of 10 $M_{\sun}$.
Models with higher upper limits would peak at younger ages, closer to
the shock; 
see MG09 for models with an IMF upper mass limit of 100 $M_{\sun}$.}

\begin{figure}
\epsscale{1.00}
\plotone{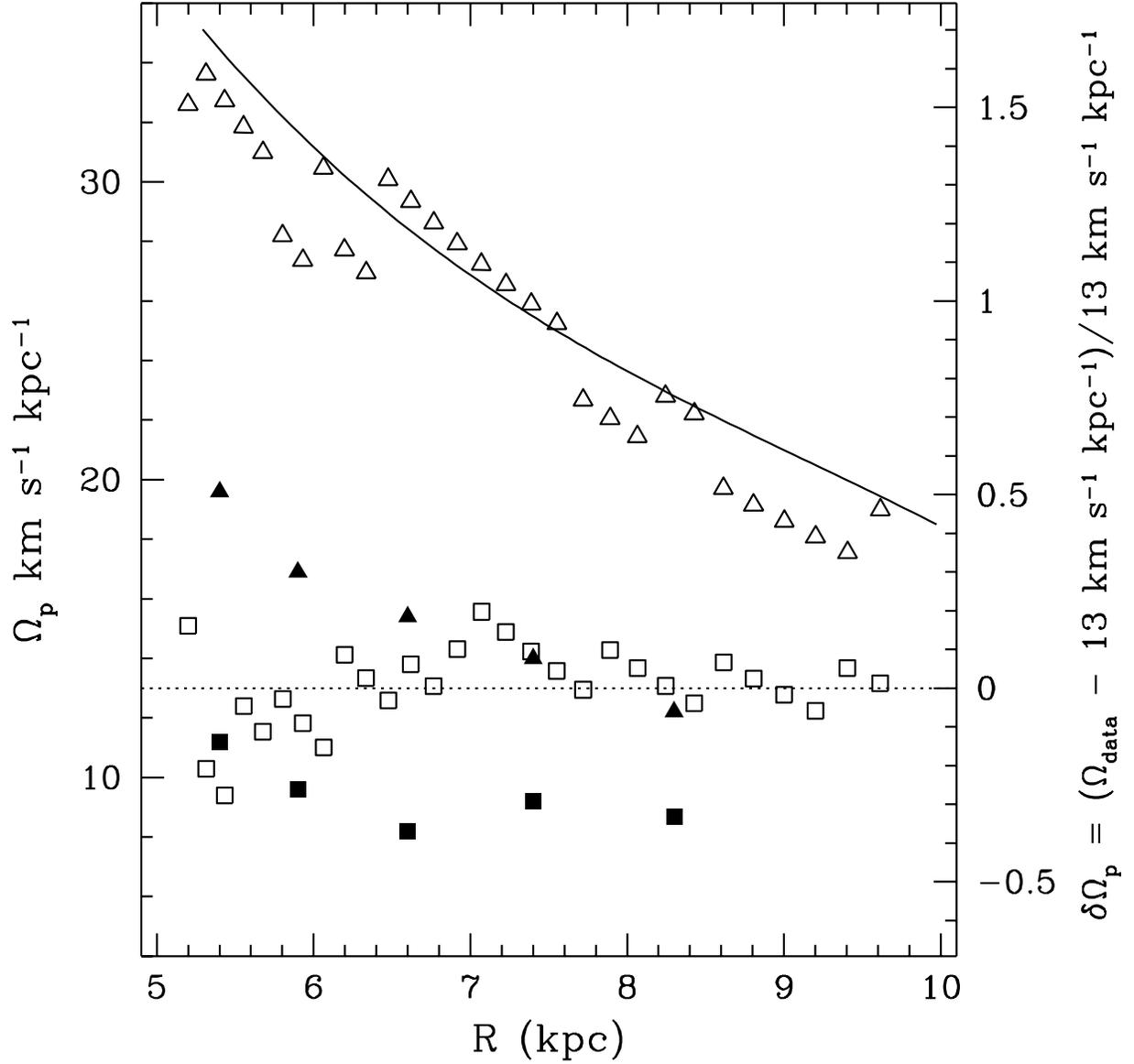}
\caption{
{\bf $\Omega_p$ measurements.}
All the analysis methods {\it assume} circular kinematics, regardless
of the actual kinematics of the input galaxy.
{\it Dotted line:} input pattern speed.
{\it Squares:} measurements, with pixel averaging,
for a synthetic galaxy with purely circular motions
obtained through the GG96 ``stretching" method ({\it solid}) and via the  
comparison of $Q$ peaks ({\it open}).
{\it Solid line:} $\Omega_p$ values derived without averaging in the case
where the  synthetic galaxy has streamline trajectories described by the
semi-analytic model.
{\it Triangles:} 
measurements, with pixel averaging, for a synthetic galaxy with
streamline trajectories described by the semi-analytic model
deduced from the GG96 method ({\it solid}) and
with the $Q$ peak method ({\it open}). 
\label{myOMEGAS}}
\end{figure}

In figure~\ref{myOMEGAS}, we illustrate the effects of
averaging. 
The dotted line is the input pattern speed.
If the synthetic galaxy has circular orbits and the galaxy is analyzed
assuming circular orbits, then one obtains the $\Omega_p$ values depicted by squares.
The solid black squares are obtained with the GG96 ``stretching" method,
whereas the comparison between model and data $Q$ maxima positions relative
to the shock yields the open symbols. Clearly,
the solid black squares are biased while the open ones are not.

On the other hand, if the synthetic galaxy has non-circular orbits but
is analyzed under the assumption of circular orbits, then one obtains
the $\Omega_{p}$ values shown as triangles.
The solid black triangles are the values obtained with the 
GG96 method;
the empty triangles are the pattern speeds found when comparing only the
positions, relative to the shock location, of model and ``observed" $Q$ peaks; and the solid line
represents the biased values measured from the semi-analytical
solutions, without averaging, if existing non-circular
motions are ignored. 
The reason for this bias towards higher measured 
pattern speeds is that real gas streamlines in a steady rotating
spiral shock turn somewhat along the arm after passing through the
shock. Consequently, stars close to the shock are slightly older than
would be expected in a circular model with the same $\Omega_p$.
The observer using a circular model would infer that the gas flow into the arms is smaller than is actually
the case, i.e., that the difference between the stellar orbital velocity and
the pattern speed is smaller than in reality; inward of corotation, this means that the observer would
overestimate the pattern speed.
Also inside corotation, the effect decreases with galactocentric radius,
as the shock strength diminishes.

The open triangles product of the peak comparison are 
much less sensitive to the competing systematics introduced by orbit averaging.
However, ironically, the original ``stretching" method (solid black triangles) 
seems much less biased towards higher values if non-circular motions
are neglected, although the dependence of the bias on galactocentric radius
is still noticeable.  
This is a fortunate turn of events, because
(1) in real images it is hard to pinpoint the location of the star formation 
onset, and (2) the radial dependence of the bias allows us to 
detect the presence of non-circular motions and to confirm the link between 
star formation and disk dynamics.
  
\section{Discussion.} 

From theory~\citep[e.g.,][]{gitt04}, it is well known that the spiral shock  
strength diminishes as one approaches corotation. 
The shock weakening implies fewer radial movements and 
more circular trajectories.  
Figures~\ref{delayt} (corotation radius, $R_{CR}$, located at $\sim 17$ kpc) and \ref{apcm}
($R_{CR} \sim 11$ kpc) support this conclusion; they also allow us to predict
that, in general, $\Omega_p$ values derived from color gradients observed inside corotation in 
real data will be too high if a circular dynamic model   
is adopted.\footnote{For some radii (e.g., $\sim$ 6 and 8.5 kpc) in the MHD simulation,
the measured $\Omega_{p}$ will be actually slightly lower ($\sim$ 1 km s$^{-1}$ kpc$^{-1}$)
than the input pattern speed of 20 km s$^{-1}$ kpc$^{-1}$.} 
For color gradients observed beyond corotation, we expect again to overestimate the 
pattern speed, but the difference
between the actual and the measured $\Omega_{p}$ will now grow with radius.
In this case, the radial velocities of the streamlines
after the shock have an outward, rather than an inward, component
\citep[see leading case in ][]{rob69}. If, however, spiral shocks are not as strong as 
inside corotation, inverse post-corotation gradients will yield unbiased
pattern speed measurements.

How would these results affect the \citet{mar09} conclusions about the 
end points of spiral patterns?
Their figure 4 displays the ratio, $R_{\mathrm{res}}/R_{\mathrm{end}}$, 
of the resonance radii (calculated from their 
pattern speed measurements and rotation curves) to the 
observed spiral end points in their deprojected, near-infrared galaxy images.
If their $\Omega_{p}$ measurements (most of them inside corotation) are systematically too high, 
then the resonance radii, and their ratio to the spiral end points would be systematically smaller than
the true values. 

In order to better quantify the impact of non-circular motions on the MG09 results, we define the expression 
$\delta \Omega_{p} =  (\Omega_{\mathrm{data}}- \Omega_{p}') / \Omega_{p}'$. Here, $\Omega_{\mathrm{data}}$
is the pattern speed obtained from the data, by comparing the observed color gradient candidates with the 
stellar models, under the assumption that stars move in purely circular orbits; $\Omega_{p}'$ is the
pattern speed that the spiral should have if it ends 
at certain resonance. For the 4:1 resonance, for example, we have:

\begin{equation}
 \Omega_{p}' \sim \frac{v_{\mathrm{rot}}}{R_{4:1}} \left(1-\frac{\sqrt{2}}{4} \right),
\end{equation}

\noindent where $v_{\mathrm{rot}}$, is the circular orbital velocity (obtained from the literature,
see MG09). And if the spiral ends at the OLR:

\begin{equation}
 \Omega_{p}' \sim \frac{v_{\mathrm{rot}}}{R_{OLR}} \left(1+\frac{\sqrt{2}}{2} \right).
\end{equation}

\begin{figure}
\epsscale{1.00}
\plotone{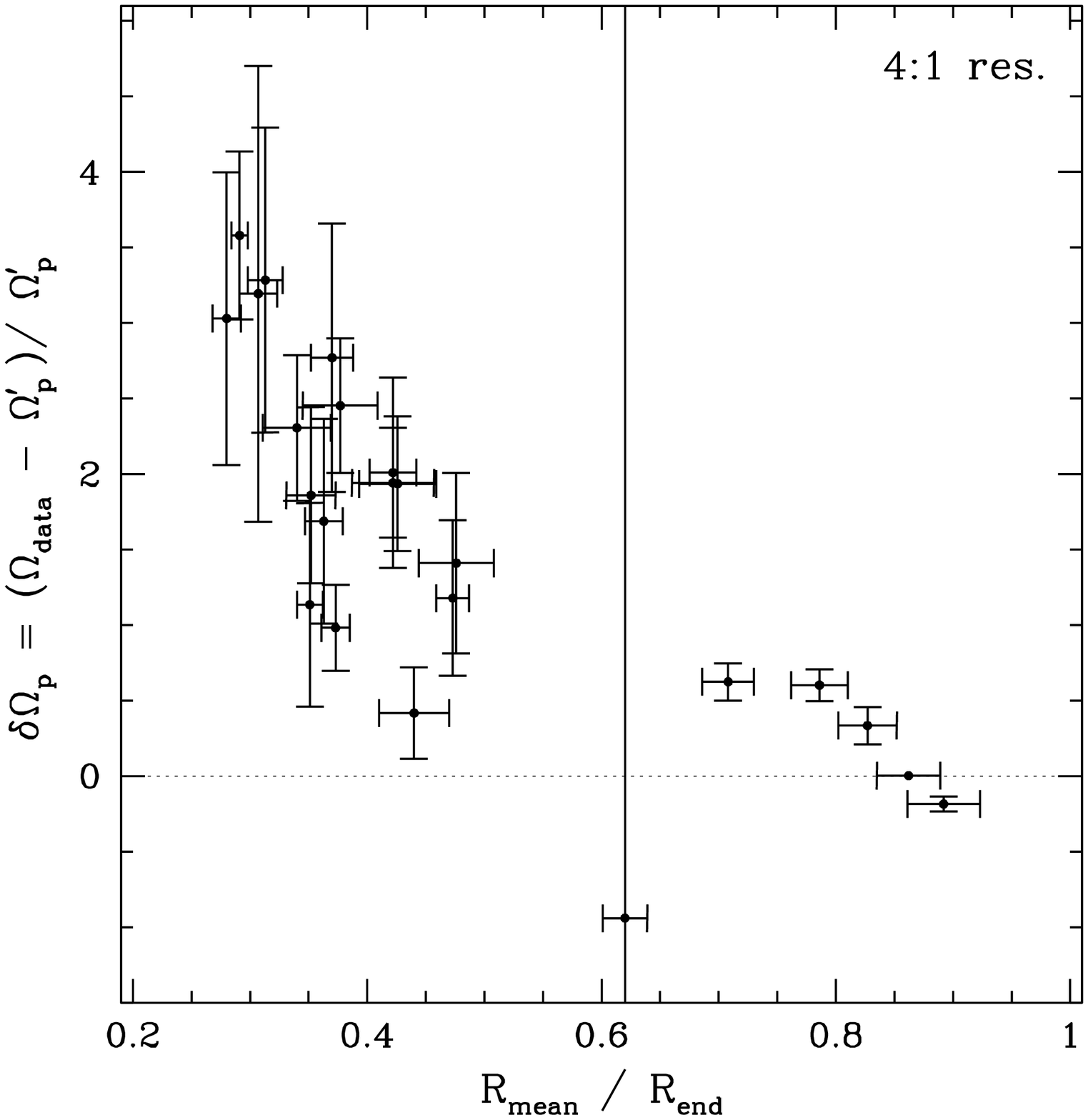}
\caption{$\delta \Omega_{p}$ (see text) vs.\ $R_{\mathrm{mean}}/R_{\mathrm{end}}$,
for the 23 regions in Table 4 of MG09. The spirals are assumed to end at the 4:1 resonance.
The dotted line indicates $\delta \Omega_{p}=0$ (i.e., no difference between the 
real and the measured pattern speeds). \label{diffOM_4_1}}
\end{figure}

\begin{figure}
\epsscale{1.00}
\plotone{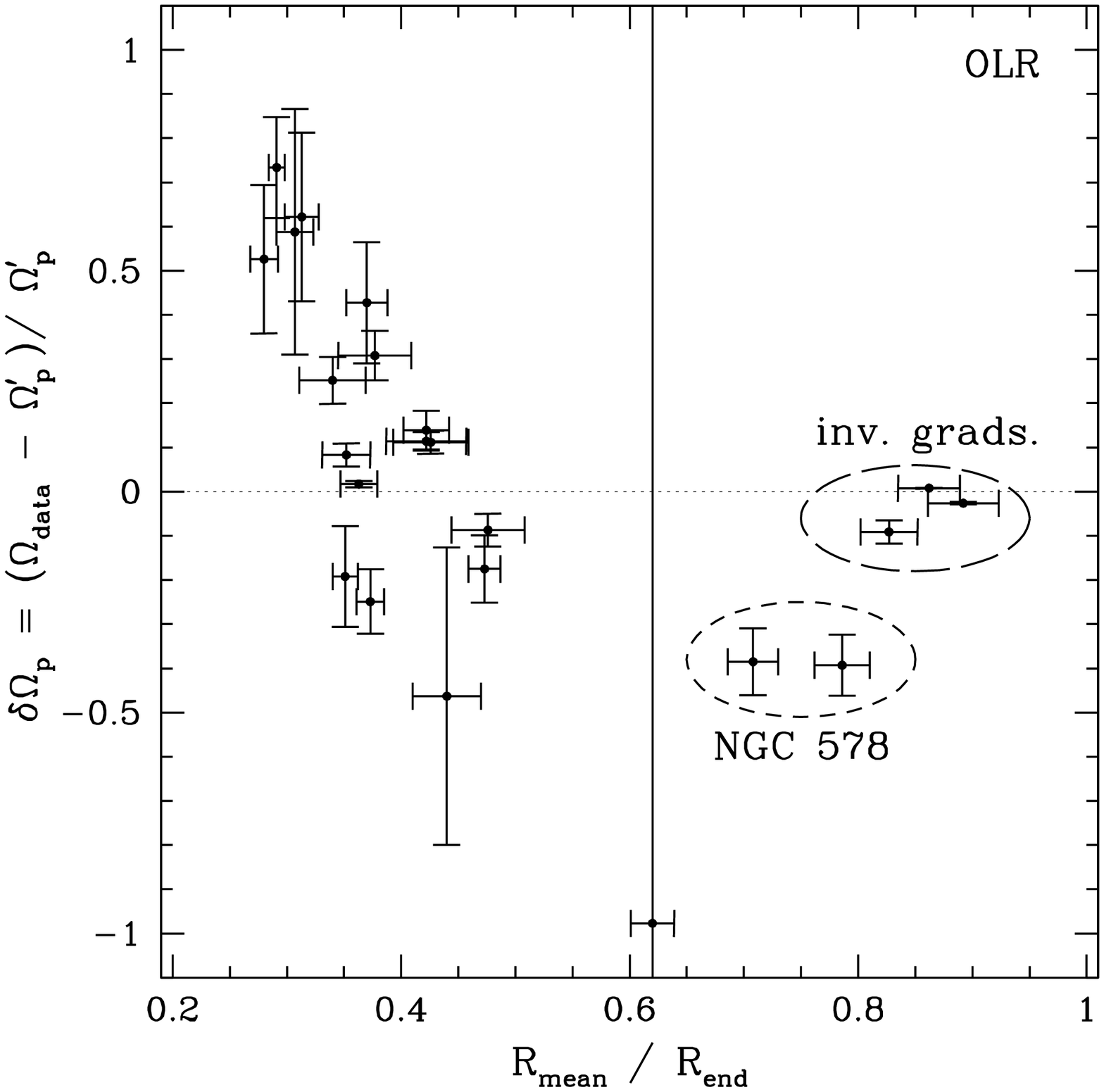}
\caption{Same as figure~\ref{diffOM_4_1} for the case where spirals are assumed to end
at the OLR. {\it Long-dashed ellipse:} Surrounds points that presumably correspond to inverse
color gradients, outside corotation. {\it Short-dashed ellipse:} Surrounds points from regions in NGC 578, 
whose spiral pattern might end at or inside corotation (MG09). \label{diffOM_OLR}}
\end{figure}

To calculate $\delta \Omega_{p}$ for the 23 regions analyzed in
MG09, we take both $\Omega_{\mathrm{data}}$ and $R_{\mathrm{end}}$ 
from their Table 4.
In figures~\ref{diffOM_4_1} and~\ref{diffOM_OLR},
we plot $\delta \Omega_{p}$ vs.\ 
$R_{\mathrm{mean}}/R_{\mathrm{end}}$, where $R_{\mathrm{mean}}$ is the mean orbital
radius of the studied region. Figure~\ref{diffOM_4_1} corresponds to the case where
the end point of the spiral pattern is fixed at the 4:1 resonance. 
\citet{conto86} state that strong spirals (Hubble types Sb or Sc) may be truncated at
this resonance.  
The $\delta \Omega_{p}$ values show the trend with radius 
expected from figures~\ref{delayt} and~\ref{apcm}.
This would imply that we are actually detecting non-circular motions
from the photometric data!

However, three discrepancies with respect to theoretical expectations are also present.
The first one is related to the systematic effects discussed in appendix~\ref{lnRcoll}. 
In spite of the tendency to actually measure values around the true pattern speed with
the method used by GG96 and MG09 (solid black triangles in figure~\ref{myOMEGAS}),  
there is a lack of points below the 
$\delta \Omega_{p} = 0$ line. The second difference is the gap around 
$R_{\mathrm{mean}}/R_{\mathrm{end}} \sim 0.6$,
where a point with huge vertical error bars is located.\footnote{
MG09 argued that this feature could be due to a star formation 
event, not triggered by spiral waves, located near corotation in NGC~1421.
}
If spirals are really truncated at the 4:1 resonance, and hence inside or near
corotation, one would expect a continuous distribution of points between
the ILR and the spiral end point (i.e., $R_{\mathrm{mean}}/R_{\mathrm{end}}=1$).
The third inconsistency concerns the actual values attained by 
$\delta \Omega_{p}$ at low $R_{\mathrm{mean}}/R_{\mathrm{end}}$.
From the solid black triangles in figure~\ref{myOMEGAS},  
the maximum expected value of $\delta \Omega_{p}$ is $\la 0.5$. The high
$\delta \Omega_{p}$ values would imply 
the presence of extremely strong shocks near the spiral arms.

Figure~\ref{diffOM_OLR}, in contrast, shows the behavior of $\delta \Omega_{p}$ for the case when
the spiral end points are fixed at the OLR. If spirals stop near this resonance,
the expected $R_{\mathrm{mean}}/R_{\mathrm{end}}$ value for corotation is $\sim 0.59$,
and the gap in the data distribution is naturally explained.
We also observe now that 
some points are located below the $\delta \Omega_{p} = 0$ line, as 
expected from figure~\ref{myOMEGAS}, owing to the  
systematic error introduced by the ``stretching" GG96 method and discussed in~\S~\ref{lnRcoll}. 
The three last points, marked with a long-dashed
ellipse, represent the ``inverse color gradients'' (i.e., located beyond corotation)
examined in MG09 (these points were treated as being inside corotation
in figure~\ref{diffOM_4_1}). The points surrounded by a short-dashed ellipse belong 
to the galaxy NGC 578. 
If NGC 578 actually ends before or at corotation (as argued by MG09), its position in figure~\ref{diffOM_4_1}
is in concordance with theoretical expectations (i.e., 
close to the $\delta \Omega_{p} = 0$ line).
On the other hand, NGC 578 may be the only
galaxy in the MG09 sample with a spiral pattern that stops at the 4:1 resonance, as
predicted by \citet{conto86}. 

\section{Conclusions.}

Under the assumption that the orbits of young stars preserve the velocity components
of the parent molecular clouds where they form, we have analyzed the
effects that non-circular motions would have on azimuthal color gradients.
Semi-analytical calculations and MHD simulations show that the spiral pattern speeds derived
from the comparison between color gradients and stellar population synthesis 
models, assuming purely circular motions, would have values systematically higher than the
real ones for regions within corotation. 
Also inside corotation, the effect would decrease with galactocentric radius.
Non-circular motions, however, would not prevent the detection of legitimate azimuthal 
color gradients in real galaxies.

Using a synthetic image, we have also analyzed the effects of image processing and pixel averaging
on the method applied in GG06 and MG09 to detect color gradients and 
derive pattern speeds. We have found that pixel averaging (due to image processing)
systematically decreases the derived $\Omega_p$, such that 
it nearly compensates for the systematic effect introduced when 
neglecting non-circular motions in the analysis. The net result is that 
the correct spiral pattern speeds and resonance locations can be obtained.
Nevertheless, a residual trend of slightly higher 
pattern speeds at lower radii can still be discerned  
({\it solid} triangles in Fig.~\ref{myOMEGAS} and Fig.~\ref{diffOM_OLR}), so that
it is possible to detect the presence of non-circular motions and 
confirm the link between star formation and disk dynamics.  

We have re-examined the results obtained by MG09. When normalizing the 
mean radii where gradients were found by the end point radii, 
in order to treat the whole sample as a single galaxy,
we were able to reproduce the trend of $\delta \Omega_{p}$ with radius expected
if non-circular motions are ignored (as these authors did). 
The size of the observed $\delta \Omega_{p}$ in the data only matches
the theoretical expectations if spiral patterns end at the
OLR, and not at the 4:1 resonance (cf.\ Fig.~\ref{myOMEGAS} with both Fig.~\ref{diffOM_4_1} and 
Fig.~\ref{diffOM_OLR}).

Future studies of azimuthal color gradients in disk galaxies by orbit calculations 
should include the effects of disk heating \citep{asi99}, and a correct
modeling of the IMF as a discrete distribution. These factors may be important,
since high mass stars may take different trajectories when compared to low mass
stars, due to the effects of dynamical friction~\citep{chandra43}. 

\acknowledgments
We thank the anonymous referee for his/her very constructive suggestions and comments that
helped us improve the presentation of this work.

E. Mart\'{\i}nez-Garc\'{\i}a acknowledges financial support from CIDA (Centro de Investigaciones de
Astronom\'{\i}a, Venezuela). Special thanks to C\'esar Brice\~no, Katherine Vivas,
Carlos Abad, Gustavo Bruzual, Gladis Magris, Jes\'us Hern\'andez, and Francisco Fuenmayor,
also at CIDA.
GCG thanks DGAPA (UNAM) grant IN-106809 and CONACyT grant 50402-F.
RAGL acknowledges DGAPA (UNAM) grant IN111007 and CONACyT grant 48589-F.

\appendix

\section{Streamline locus determination from semi-analytical solutions.}
\label{appsemianalyt}

\begin{figure}
\epsscale{1.00}
\plotone{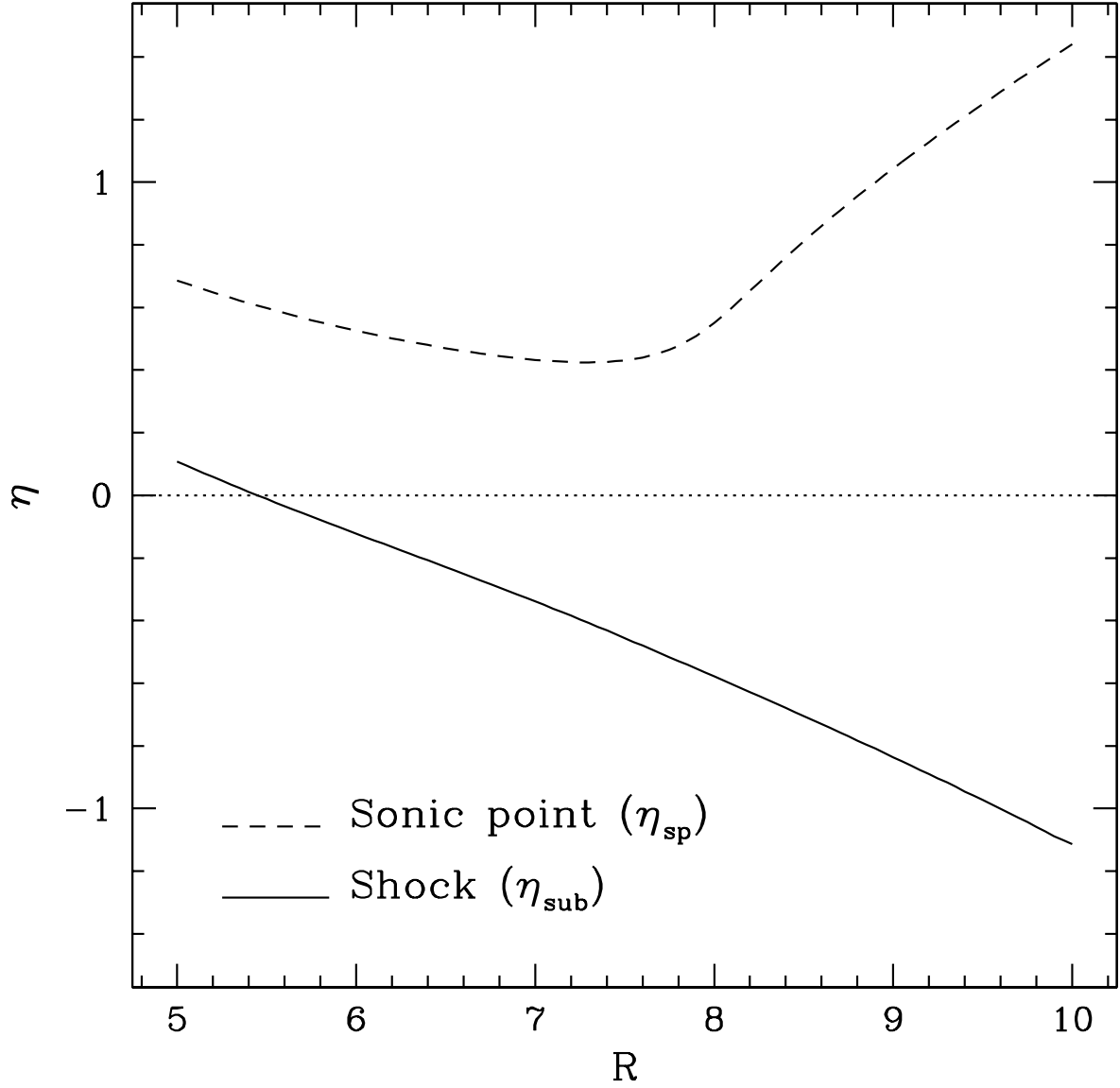}
\caption{Sonic point and shock positions at different radii. {\it Dashed line:} $\eta_{sp}$, sonic point
         position relative to the potential minimum ({\it dotted line}, $\eta=0$);
         {\it solid line:} $\eta_{sub}$, position of the 
          spiral shock relative to the potential minimum.
         \label{etasp}}
\end{figure}

The differential equations needed to obtain the density and velocity components,
product of spiral shocks, are fully explained in \citet{rob69}, \citet{shu72},
\citet{shu73}, and \citet{gitt04}.\footnote{See also \citet{fuj68}, and \citet{ishi84}.} 
The independent parameters used in this investigation for the solutions are:
number of arms, $m=2$; spiral arms pitch angle $i$ such that $\sin{i}=0.1$; angular speed of the spiral pattern,
$\Omega_{p}=13$ km s$^{-1}$ kpc$^{-1}$; effective speed of sound, $a=8$ km s$^{-1}$;
amplitude ratio of the perturbed spiral field to the axisymmetric field, $F=0.05$;
rotation velocity of material,\footnote{
A generic flat rotation curve is adopted, such that 
the epicyclic frequency $\kappa=\sqrt{2} \Omega$. We also assume that $F\sim$constant
for the selected range of radii (5-10 kpc).}
$v_{rot}=220$ km s$^{-1}$.
In figure~\ref{etasp}, we show the positions of the sonic point and the 
spiral shock relative to the potential minimum ($\eta=0$), with the parameters described above.
As already noticed by \citet{tos73} and \citet{gitt04}, the shock front moves away from
the potential minimum as the radius increases. In this case, for $R\sim5$ kpc, the shock
front is slightly ahead of the potential minimum. These relative positions may change
for another set of parameters. 

The solutions are obtained in curvilinear coordinates $\eta$ (perpendicular to the spiral equipotential curves)
and $\xi$ (parallel to the spiral equipotential curves).
We adopt the $\eta$ and $\xi$ definitions from \citet{shu73} and \citet{gitt04}.
These definitions differ from those used in \citet{rob69} and \citet{shu72} by a multiplicative factor
$m/\sin{i}$, and by an additive constant, chosen such that $\eta=0$ determines
the location of the potential minumum. The corresponding expressions in terms of the radius, $R$, and the
angular coordinate in the rotating frame of the spiral pattern, $\Theta=\theta- \Omega_{p}t$, are as follows:

\begin{equation}
   \eta = \frac{m}{\tan{i}} \ln\left({\frac{R}{R_{a}}}\right) + m (\theta- \Omega_{p}t) + \pi,
\end{equation}

\begin{equation}
    \xi = -m \ln\left({\frac{R}{R_{a}}}\right) + \frac{m}{\tan{i}} (\theta- \Omega_{p}t).
\end{equation}

The $\eta$ coordinate is obtained from the shock solution, together with the velocity components
$u_{\xi} = u_{\xi0} + u_{\xi1}$, and $u_{\eta} = u_{\eta0} + u_{\eta1}$. The subscript 0 labels 
the unperturbed velocity (i.e., in absence of a spiral perturbation), and the subscript 1, the
perturbation due to the spiral gravitational field.
The $\xi$ coordinate is obtained by solving the equation:

\begin{equation}
    \frac{d\xi}{d\eta}=\frac{u_{\xi}}{u_{\eta}},
\end{equation}

\noindent that can also be expressed as: 
\begin{equation}
    \int^{\eta_{stream}}_{\eta_{sub}} \left(\frac{d\xi}{d\eta}\right) d\eta = \xi_{stream} - \xi_{sub}.
\end{equation}

\noindent
The shock on the subsonic branch occurs at $\eta_{sub}$ and $\xi_{sub}$.
$\eta_{stream}$ and $\xi_{stream}$ are coordinate values at any point
along the streamline; their maximum possible values are
$\eta_{sup}$ and $\xi_{sup}$, i.e., the point at which the shock occurs on the supersonic branch.\footnote{
$\eta_{sup}-\eta_{sub}=2\pi$.
In theory, $\xi_{sup}-\xi_{sub}=\frac{2\pi}{\tan{i}}$ but, due to a slight
nonclosure of the streamlines with radius, the numerical values may differ \citep[see][]{shu72}.}

In order to obtain $\xi_{sub}$, we proceed in the way described below.
We define the quantity:
\begin{equation}
    x = R_{0} \left( { \frac {\int^{\Theta_{sup}'}_{\Theta_{sub}'} \left(\frac{R}{R_{a}}\right)' d\Theta'}
                                {\Theta_{sup}' - \Theta_{sub}'} } \right)^{-1},
\end{equation}
where $\left(\frac{R}{R_{a}}\right)'$ and $\Theta'$ are the $\left(\frac{R}{R_{a}}\right)$ and $\Theta$
values obtained when $\xi_{sub}=0$ and $R_{a}=1$. $R_{0}$ is the average radius of the streamline.

We use the $x$ value to get:

\begin{equation}
    \xi_{sub} = - \left(m+\frac{m}{\tan^{2}{i}}\right) \ln{x}.
\end{equation}

\noindent
With this definition, the mean value of $\left(\frac{R}{R_{a}}\right)$ will be $R_{0}$.

The time elapsed since the shock, $t$, is obtained by integrating the velocity along
the $\Theta$ direction:

\begin{equation}
   t = \int^{\Theta_{stream}}_{\Theta_{sub}} \frac{R d\Theta}{u_{\Theta}},
\end{equation}

\noindent
where $t=0$ when $\Theta_{stream}=\Theta_{sub}$, and

\begin{equation}
   u_{\Theta} = \frac{u_{\eta} + u_{\xi}\cot{i}} {\cot{i}\cos{i} + \sin{i}}.
\end{equation}


\end{document}